# Negative Refraction and Left-handed electromagnetism in a Microwave Photonic Crystal

P. V. Parimi[1], W. T. Lu[1], P. Vodo[1], J. Sokoloff[1], J. S. Derov[2], and S. Sridhar[1]*

[1]Department of Physics and Electronic Materials Research Institute, Northeastern University,

360 Huntington Avenue, Boston, MA 02115.

[2] AFRL/SNHA, 80 Scott Drive, Hanscom AFB, MA 01731

We demonstrate negative refraction of microwaves in a metallic photonic crystal prism. The spectral response of the photonic crystal prism, which manifests both positive and negative refraction, is in complete agreement with band-structure calculations and numerical simulations. The validity of Snell's law with negative refractive index is confirmed experimentally and theoretically. The negative refraction observed corresponds to left-handed electromagnetism that arises due to the dispersion characteristics of waves in a periodic medium. This mechanism for negative refraction is different from that in metamaterials.

78.20.Ci, 42.70.Qs, 41.20.Jb

The optical properties of materials that are transparent to electromagnetic (EM) waves can be characterized by an index of refraction $n$. Given the direction of the incident beam $\theta_1$ at the interface of vacuum and the material, the direction $\theta_2$ of the outgoing beam can be determined using Snell's law $\sin\theta_1 = n\sin\theta_2$. All naturally available materials possess a positive refractive index $n > 0$. It was observed recently that in certain composite metamaterials EM waves bend negatively[1] and consequently a negative index of refraction $n < 0$ can be assigned to such materials without violating Maxwell's equations. For homogeneous media, negative refraction ($\theta_2 < 0, \theta_1 > 0$) necessarily requires that $n < 0$. This negative bending allows considerable control over light propagation and opens the door for new approaches to a variety of applications from microwave to optical frequencies.

Negative index media exhibit some unusual propagation characteristics of EM waves. The most striking property is that of Left-Handed Electromagnetism (LHE) since the electromagnetic fields $\vec{E}$ and $\vec{H}$, and the wave vector $\vec{k}$ form a left-handed triplet. Consequently the energy flow represented by the Poynting vector $\vec{S} = \vec{E} \times \vec{H}$ is anti-parallel to the wave vector $\vec{k}$, so that $\vec{S} \cdot \vec{k} < 0$. In contrast for conventional $n > 0$ materials $\vec{E}, \vec{H}, \vec{k}$ form a right-handed triplet corresponding to Right-Handed Electromagnetism (RHE), and $\vec{S} \cdot \vec{k} > 0$.

A material possessing simultaneously negative permittivity $\varepsilon < 0$ and permeability $\mu < 0$ can be shown to necessarily have $n \equiv \sqrt{\varepsilon}\sqrt{\mu} < 0$[2,3]. Recently, negative refraction was demonstrated in a quasi-homogeneous metamaterial [1,4,5] with unit cell dimensions less than the wavelength, consisting of interleaved arrays of wires ($\varepsilon < 0$) and split ring resonators ($\mu < 0$). However these materials are highly absorptive, and unlikely to be scaled to three dimensions or to smaller sizes suitable for applications at optical frequencies [1,6,7].

It has since been proposed that negative refraction can be achieved in photonic crystals (PC), which are inhomogeneous periodic media with lattice constant comparable to the wavelength [8,9]. A PC is an artificial structure, usually made of a dielectric or metal, designed to control photons similar to the way a solid state crystal controls electrons. Locally both $\varepsilon, \mu > 0$ everywhere in the PC. The physical principles that allow negative refraction in the PC arise from the dispersion characteristics of wave propagation in a periodic medium, and are very different from that of the metamaterial in Ref. [1]. In this Letter we present experimental evidence of negative refraction of microwaves in a metallic photonic crystal. Parallel theoretical and numerical investigations of the band structure and simulations of wave propagation through the PC prism are compared with the experimental results. Exceptionally good agreement is found between the experiments, band structure calculations and wave refraction simulations.

The microwave photonic crystal fabricated in the present work is an array of cylindrical copper rods of height 1.26 cm and radius 0.63 cm forming a triangular lattice. The ratio of the radius $r$ to lattice constant $a$ is $r/a = 0.2$. Refraction measurements are carried out in a parallel plate waveguide made of a pair of metallic plates. The excitation in the parallel plate waveguide is transverse magnetic (TM) mode up to 12 GHz such that the electric field $\vec{E}$ is parallel to the rod axis. Microwave absorbers are placed on both sides to avoid spurious reflections and to collimate the propagating beam of width 9 cm, which is incident normally on a right angle prism of PC (see Fig. 1(a)). On the far side, a dipole antenna attached to an X-Y robot maps the electric field $\vec{E}$. Accurate angles of refraction are obtained by fitting the emerging wave front with a plane wave, with the refraction angle $\theta_2$ as a fit parameter.

The refraction experiment is validated by data on a polystyrene prism having similar dimensions as that of the PC prisms. In Fig. 1(f) the direction of the emerging beam can be clearly seen at an angle $\theta_2 = +52.2^0$ from the normal to the surface of refraction, corresponding to a positive refractive index $n = 1.58$ ($n \sin \theta_1 = \sin \theta_2$) for an incident angle $\theta_1 = 30^0$.

The measurements on the triangular lattice PC were carried out with the incident wave vector $\vec{k}_i$ along directions $\Gamma \to K$ and $\Gamma \to M$ of the first Brillion zone (these directions in real space and in reciprocal wave-vector space are shown in Fig.3). The angles of incidence, $\theta_{1K} = 30°$ for $\Gamma \to K$ and $\theta_{1M} = 60°$ for $\Gamma \to M$, are chosen in such a way that the periodicity on the surface of refraction is minimum to prevent higher order Bragg diffraction[5]. Fig. 1(c) illustrates the negatively refracted wave front with an angle of refraction $\theta_{2K} = 11.5°$ for f = 9.77 GHz with incidence along $\Gamma \to K$. Using Snell's law, $n_{eff} \sin \theta_{1K} = \sin \theta_{2K}$ with $\theta_{1k} = 30°$, $\theta_{2K} = -11.5°$, we obtain an *effective* refractive index of $n_{eff} = -0.4$ at this frequency. A second wave front can also be seen emerging from the top edge of the PC. We attribute this wave front to the edge effect due to the finite sample size. The negative refraction reported in the present photonic crystal has been demonstrated for the first time in an inhomogeneous system[10]. Fig 1e shows positive refraction at f = 6.62 GHz. In Fig. 1d an illustrative example of negative refraction at 10.4 GHz for $\Gamma \to M$ ($\theta_{1M} = 60°$) is shown.

An understanding of negative refraction and its relation to left-handed behavior of electromagnetic waves in a photonic crystal can be achieved by examining the band structure of an infinite PC. We have calculated the band structure employing standard plane wave expansion methods using over 2000 plane waves[11]. The 2D band structure for a triangular lattice PC with $r/a = 0.2$ is shown in Fig. 2(a).

For a plane wave with wave vector $\vec{k}_i$ and frequency $\omega$ incident normal to an air-PC interface, the wave vector $\vec{k}_f$ inside the PC is parallel or anti-parallel to $\vec{k}_i$ as determined by the band structure. If $d\omega/d|\vec{k}_f| > 0$, $\vec{k}_f$ is parallel to $\vec{k}_i$ and consequently the EM field in PC is right-handed (RHE). Otherwise, $\vec{k}_f$ is anti-parallel to $\vec{k}_i$ and the EM field in the PC is left-handed (LHE). For a general case the phase and group velocities in a medium are $\vec{v}_p = (c/|n_p|)\hat{k}_f$ with $\hat{k}_f = \vec{k}_f /|\vec{k}_f|$ and $\vec{v}_g = \nabla_{\vec{k}}\omega$. It can be proved analytically that the direction of group velocity $\vec{v}_g$ in an infinite PC coincides with that of the energy flow $\vec{S}$ [12]. An effective refractive index can be defined $n_p = \text{sgn}(\vec{v}_g \cdot \vec{k}_f) \frac{c|\vec{k}_f|}{\omega}$ and calculated from the band structure [13]. The sign of $n_p$ is determined form the behavior of the equi-frequency surfaces (EFS). EFS plots for the first and second bands of the triangular lattice are shown in Fig.3(a). The center in the plots corresponds to the center of the first Brillouin zone. The EFS that move outwards from the center with increasing frequency correspond to RHE with $\vec{v}_g \cdot \vec{k}_f > 0$ and inward moving surfaces correspond to LHE with $\vec{v}_g \cdot \vec{k}_f < 0$. In the case of LHE (RHE) conservation of $\vec{k}_f$ component along the surface of refraction would result in negative (positive) refraction. The resulting refractive index $n_p$ determined from the band structure and EFS for a beam incident along both $\Gamma \to K$ (dashed line) and $\Gamma \to M$ (solid line) is shown in Fig. 2(c). Note that negative refraction is predicted for regions in the 2nd and 3rd bands and positive refraction in the 1st and 4th bands.

In the following we describe the salient features of the experimental results and comparison to band structure;

(I) In the first band between 6.2 – 7.7 GHz the EFS move outward with increasing frequency, so that $n_p > 0$ corresponding to RHE with $\vec{v}_g \cdot \vec{k}_f > 0$ (i.e. $\vec{v}_g$ parallel to $\vec{k}_f$). The representative field map at f = 6.62 GHz in Fig. 1(e) confirms the positive refraction expected. The measured $n_{eff}$ are in good agreement with the theoretical calculations (Fig. 2(c)).

(II) In the second band between 7.7 – 11 GHz, the EFS move inward with increasing frequency, consistent with $n_p < 0$ corresponding to LHE with $\vec{v}_g \cdot \vec{k}_f < 0$ ($\vec{v}_g$ anti-parallel to $\vec{k}_f$). The illustrative field map in Fig. 1(c) at f = 9.77 GHz shows the emerging wave front in the negative direction. The experimental results of refraction for the incident beam along both $\Gamma \to K$ ( $\Delta$ in Fig 2(c)) and $\Gamma \to M$ ( $*$ ) are in excellent agreement with the band structure calculations (dashed and solid lines in Fig. 2(c)).

(III) In certain frequency ranges in which the EFS is circular and frequency is not so high, the 1st order Bragg diffraction is very weak. For the 0th order Bragg peak, $n_p$ is angle independent and consequently Snell's law is applicable. The index of refraction $n_p$ (see Fig 2(c)) in the region 8.8 -11 GHz determined from the experimental wave field scans for different angles of incidence, viz., $\theta_{1K} = 30°$ for $\Gamma \to K$ and $\theta_{1M} = 60°$ for $\Gamma \to M$, is nearly angle independent due to the circular nature of the EFS, confirming the validity of Snell's law in this frequency region. *Thus we have confirmed the validity of Snell's law both experimentally and theoretically* ( Fig.2(c)). A noteworthy point is that the strong contrast in the metallic PC leads to near circular EFS (Fig 3(a) ) and thus results in negative refraction in a wider frequency range than that of a dielectric PC [6,8,9].

(IV) The band cutoff at 6.2 GHz , transmission between 6.2 – 11.1 GHz and band gap region between 11.1-11.3 GHz, all of which are observed in the transmitted spectrum shown in Fig. 2(b), are in excellent agreement with the band structure calculations shown in Fig. 2(a).

Direct numerical simulations (see Fig. 3) of wave refraction were also carried out and are in good agreement with the experimental results and band-structure calculations. The simulation is done using a Green's function boundary wall approach originally developed for hard wall potentials in quantum mechanics [14]. As shown in Fig. 3, the simulation results in negative refraction for $f$ = 9.7 GHz in the 2nd band, and positive refraction for $f$ = 6.6 GHz in the 1st band, both in agreement with the experiment (Fig. 1).

In conclusion we have experimentally demonstrated negative refraction in a new class of materials, metallic photonic crystal. A major feature of the present work is the extraordinary level of control exemplified by the convergence between the experimental data, band structure calculations and simulations. This means that a variety of tailor-made structures are feasible that can be designed and constructed. There are numerous possibilities opened up by the present results. For many applications such as imaging, one requires index matching between the negative index material and surroundings (relative index $n = -1$) accompanied by negligible losses. These requirements are more easily met with PC than with composite negative index metamaterials. Metallic PC offers the additional advantages of the highest dielectric contrast, low attenuation and the possibility of focusing, which are evident from the present data. Furthermore the microwave PC can be easily scaled to 3 dimensions[15], and to optical frequencies [16], which is highly unlikely with composite negative index metamaterials [1,6]. Thus the advantages of negative refraction and left-handed electromagnetism that have been proposed recently, such as imaging by flat lenses [17], beam steerers, couplers and others, as well as some entirely new possibilities, are feasible with photonic crystals from microwave to optical frequencies.

Work supported by the National Science Foundation and the Air Force Research Laboratories.

*Electronic address: s.sridhar@neu.edu

Fig.1. [Color] (a) Schematic diagram of microwave refraction experimental setup ( not to scale). (b) Propagation vectors for positive and negative refraction. (c, d, e &f) Microwave electric field maps in the far field region. (c) Negative and (e) positive refraction the by metallic PC prism for the incident beam along $\Gamma - K$ forming an angle of incidence $30°$ with the refraction surface. WF = wave front with respect to refracting surface. (d) Negative refection for the incident beam along $\Gamma - M$ forming an angle of incidence $60°$. (f) Positive refraction by a polystyrene prism. In all the field maps approximate area of each field map is 43X40 cm$^2$, the PC prisms and incident beams are schematic and do not correspond to the actual sizes used.

Fig 2. (a) Band structure computed for the triangular lattice metallic PC with $r/a = 0.2$. (b) Microwave transmission amplitude $|S_{21}|$ vs. frequency f (GHz) on the far side. Note the band cutoff below 6.2 GHz, transmission between 6.2 – 11.1 GHz and band gap region 11.1-11.3 GHz, are in excellent agreement with Fig.2 (a). (c) Refractive index $n_p$ determined from the experimental results for a beam incident along $\Gamma \rightarrow K$ ($\Delta$) and $\Gamma \rightarrow M$ ($*$). For $n_p < 0$ ($n_p > 0$), the electromagnetism is left-handed (right-handed), and the refraction angle is negative (positive). The close match between the experiment ($\Delta$ and $*$) and theory (dashed and solid lines) is evident and shows agreement with Snell's law.

Fig. 3. [Color] (a) The EFS for the 1$^{st}$ and 2$^{nd}$ bands. The centers of the hexagons represent the centers of the first Brillouin zones in the respective bands. Blue (red) color represents lower (higher) frequency. (b & c) Simulations of wave refraction showing the wave front emerging from a metallic PC: (b) Positive refraction at 6.6 GHz and (c) negative refraction at 9.7 GHz. The PC used in the simulations has the same size as that used in the experiment. The electric field is plotted as $E_z^{1/3}$ for better visibility. The agreement with Fig. 1 (e) and (c) is evident.

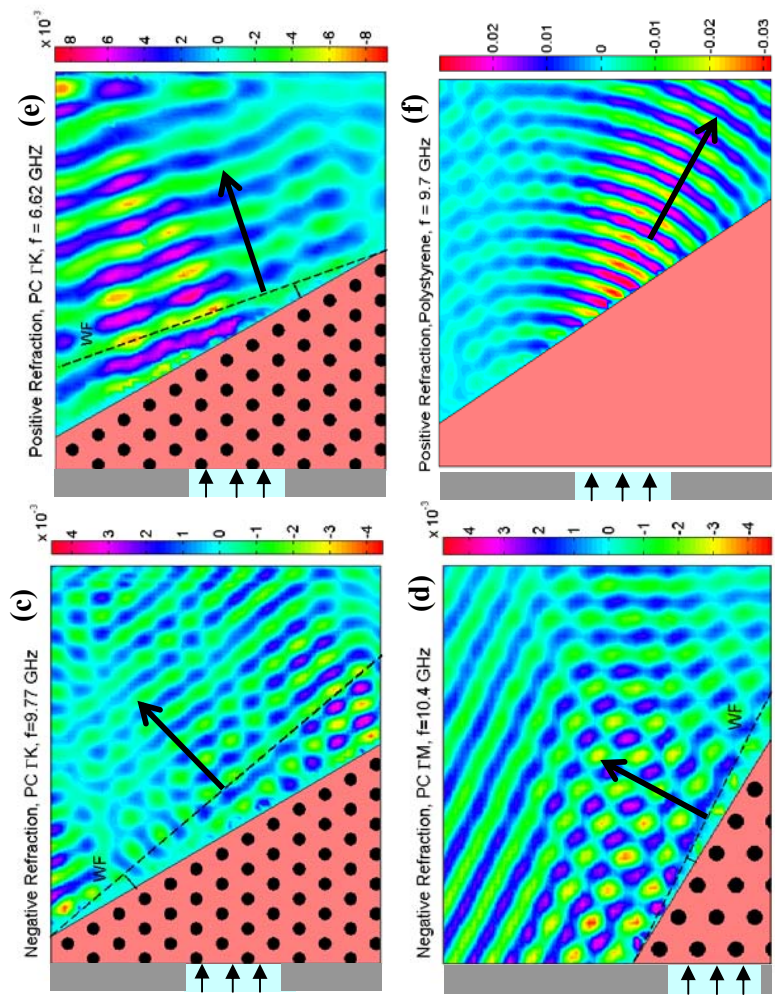
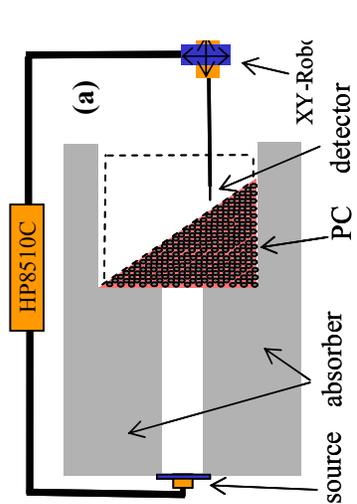
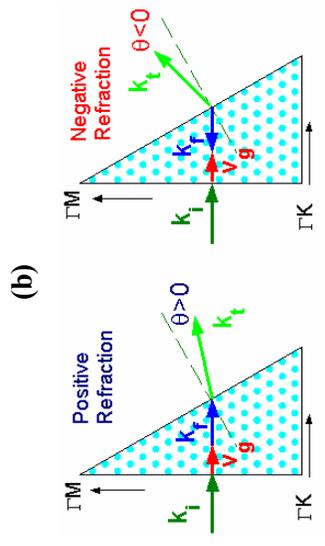

Fig. 1 P.V. Parimi et al

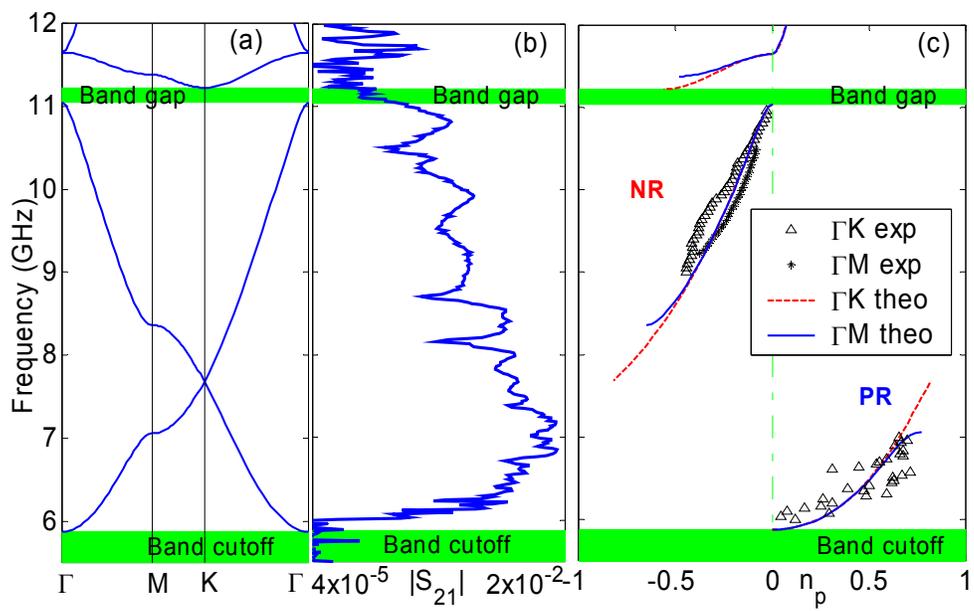

Fig. 2 – P. V. Parimi et al.

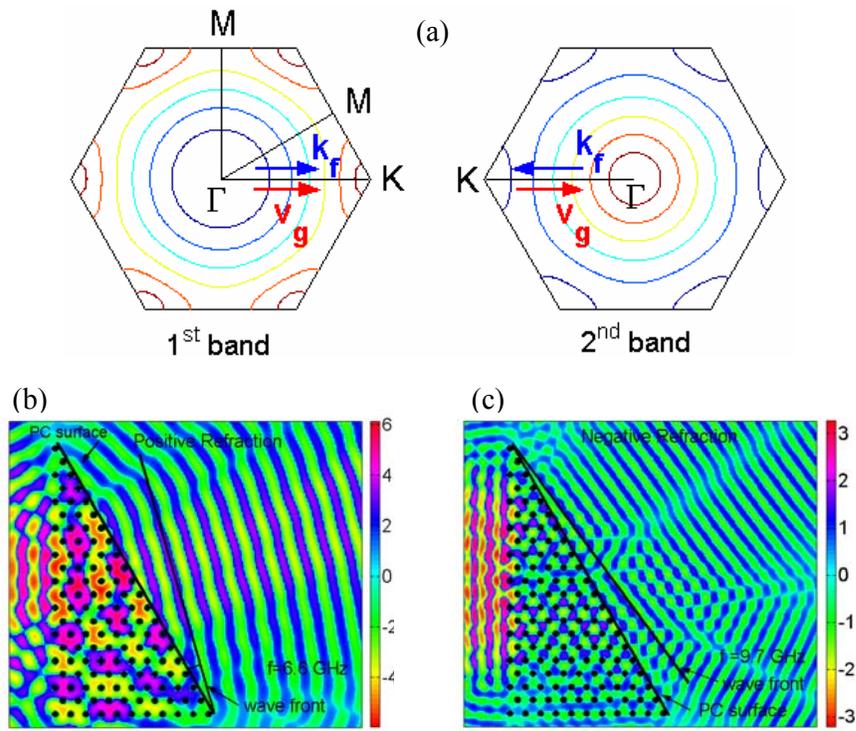

Fig 3. P.V. Parimi et al